\documentclass[review]{elsarticle}
\usepackage{color}
\usepackage{lineno,hyperref}
\modulolinenumbers[5]
\usepackage{amssymb}
\usepackage{amsmath}
\usepackage{xspace}

\newcommand{\Jpsi}{\ensuremath{{\rm J}/\psi}\xspace}
\newcommand{\Rz}{\ensuremath{\rho^0}\xspace}

\newcommand{\fivenn}       {$\sqrt{s_{\mathrm{NN}}}~=~5.02$~Te\kern-.1emV\xspace}

\newcommand{\gPb}{\ensuremath{\gamma \mathrm{Pb}}\xspace}

\newcommand{\mant}         {\ensuremath{|t|}\xspace}

\journal{}

\begin{document}

\newpage
\begin{frontmatter}

\title{
Incoherent \Jpsi production at large \mant identifies the onset of saturation at the LHC \\
}

\author[CVUT]{J. Cepila}
\author[CVUT]{J. G. Contreras}
\author[CVUT]{M. Matas}
\ead{marek.matas@fjfi.cvut.cz}

\author[CVUT]{A. Ridzikova}
\address[CVUT]{Faculty of Nuclear Sciences and Physical Engineering,
Czech Technical University in Prague, Czech Republic}


\begin{abstract}
We predict that the onset of gluon saturation can be uniquely identified using incoherent \Jpsi production in Pb--Pb collisions at currently accessible energies of the LHC. The diffractive incoherent photo-production of a \Jpsi vector meson off a hadron provides information on the partonic structure of the hadron. Within the Good-Walker approach it specifically measures the variance over possible target configurations of the hadronic colour field. For this process then, gluon saturation sets in when the cross section reaches a maximum, as a function of the centre-of-mass energy of the photon-hadron system ($W$), and then decreases.
We benchmark the energy-dependent hot-spot model against data from HERA and the LHC and demonstrate a good description of the available data. We show that the study of the energy dependence of the incoherent production of \Jpsi allows us to pinpoint the onset of saturation effects by selecting the region of  Mandelstam-$t$ around 1 GeV$^2$ where the contribution of hot spots is dominant. We predict
the onset of saturation in a Pb target to occur for $W$ around a few hundred GeV. This can be measured with current data in ultra-peripheral Pb--Pb collisions at the LHC.
\end{abstract}


\end{frontmatter}

\section{Introduction
\label{sec:intro}}

The structure of hadrons undergoing a hard interaction is described within perturbative quantum chromodynamics (QCD) in terms of quarks and gluons, collectively denoted as partons. At high energies, equivalently small Bjorken-$x$, this structure is expected to  transition from a dilute to a saturated regime where, even though the interaction is hard, the partons interact with each other and reach a state of dynamic equilibrium between production and annihilation processes~\cite{Gribov:1983ivg,Mueller:1989st,Albacete:2014fwa}. 

According to the precise inclusive measurements from HERA~\cite{H1:2015ubc} the structure of the proton at small Bjorken-$x$ is dominated by gluons, which motivates the use of observables highly sensitive to the gluonic distribution of hadrons to search for saturation effects. This is the case for the diffractive photoproduction of vector mesons, as pointed out already in Ref.~\cite{Ryskin:1992ui}. Nowadays, it is common to study this process within the colour dipole picture where the process factorises in the splitting of the photon into a quark-antiquark dipole, the interaction of the dipole with the hadron, and the formation of the vector meson out of the scattered dipole~\cite{Kowalski:2006hc}. This process has been extensively studied at HERA~\cite{Newman:2013ada} and LHC~\cite{Contreras:2015dqa,Klein:2019qfb}. It is also an important component of the physics program of future facilities like the EIC~\cite{Accardi:2012qut} or the LHeC~\cite{LHeCStudyGroup:2012zhm}.

An attractive formalism to study diffractive processes is the Good-Walker approach~\cite{Good:1960ba,Miettinen:1978jb}. In this framework, the vector meson production off the coherent target colour field samples the average---over all possible configurations---of the gluon distribution, while the incoherent production samples the variance over the configurations. 
This formalism was applied to diffractive vector meson production for the first time in Ref.~\cite{Mantysaari:2016ykx} where it was used to describe HERA data on \Jpsi photo-production, at fixed centre-of-mass energy of the photon-proton system, in an approach that models the transverse structure of the proton as three hot spots whose position and colour charges fluctuate event-by-event. Since then, several other models based on hot spots have been proposed to study the structure of hadrons, e.g. Refs.~\cite{Traini:2018hxd,Kumar:2021zbn,Demirci:2022wuy}; for a review see Ref.~\cite{Mantysaari:2020axf}.

We proposed a variation of the hot-spot model, where the number of hot spots grows with energy~\cite{Cepila:2016uku}. In this case, the incoherent production of \Jpsi off protons provide a striking signature of saturation where the cross section, studied as a function of the centre-of-mass energy of the photon--proton system, reaches a maximum and then decreases. Within the model, this happens when the transverse area of the protons has so many hot spots that all configurations start to resemble each other in a process reminiscent of percolation~\cite{Armesto:1996kt}. Later on, our model was extended to study production off nuclear targets~\cite{Cepila:2017nef}, the dependence on the mass of the vector meson~\cite{Cepila:2018zky}, and to predict the cross section for the electro-production of vector mesons in future facilities~\cite{Bendova:2018bbb,Krelina:2019gee}.

In this Letter, we identify a new observable to determine the onset of saturation at LHC. We propose to measure the incoherent production of vector mesons off nuclear targets as a function of energy at different values of the Mandelstam-$t$ variable, which is related through a Fourier transform to the distribution of colour charges in the impact-parameter plane. The key insight is that scanning the energy behaviour in specific \mant ranges samples fluctuations of different transverse sizes and allows for the isolation of the contribution of hot spots where one expects saturation effects to set in. This will then allow for a unique identification of the presence of saturation effects at energies currently  available at LHC.  This Letter is organised as follows. A brief review of the formalism is presented in Sec.~\ref{sec:Overview}. The prediction for the onset of saturation at the LHC using this observable is presented in Sec.~\ref{sec:Results}. Our findings are discussed in Sec.~\ref{sec:Discussion}. We summarise our results and present an outlook in Sec.~\ref{sec:Summary}.

\section{Brief overview of the energy-dependant hot-spot model
\label{sec:Overview}}

In the Good-Walker approach the coherent  cross section  for the diffractive photo-production of a  vector meson V off a  hadron target H  is given by
\begin{equation}
\frac{\mathrm{d}\sigma^{\gamma^*{\rm H} \rightarrow {\rm VH}}}{\mathrm{d}|t|} \bigg| _{\rm T,L} = \frac{\left(R_g ^{\rm T,L}\right)^2}{16\pi} | \langle \mathcal{A}_{\rm T,L} \rangle |^2,
\label{VM-cs-diff-excl}
\end{equation}
with the scattering amplitude
\begin{equation}
\mathcal{A}_{\rm T,L}(x,Q^2,\vec{\Delta}) = i \int \mathrm{d}\vec{r} \int \limits_0^1 \frac{\mathrm{d}z}{4\pi} \int \mathrm{d}\vec{b} |\Psi_{\rm V}^* \Psi_{\gamma^*}|_{\rm T,L} \exp \left[ -i\left( \vec{b} - (\frac{1}{2}-z)\vec{r} \right)\vec{\Delta} \right] \frac{\mathrm{d}\sigma^{\rm dip}_{\rm H}}{\mathrm{d} \vec{b}}.
\label{VM-amplitude}
\end{equation}

 The expression for the incoherent cross section is related to the variance of scattering amplitude as
\begin{equation}
\frac{\mathrm{d}\sigma^{\gamma^*p \rightarrow {\rm V}Y}}{\mathrm{d}|t|} \bigg| _{\rm T,L} = \frac{\left(R_g ^{T,L}\right)^2}{16\pi} \left( \langle |\mathcal{A}_{\rm T,L}|^2 \rangle - | \langle \mathcal{A}_{\rm T,L}  \rangle|^2 \right).
\label{VM-cs-diff-disoc}
\end{equation}

In the previous formulas, T and L stand for the transverse and longitudinal contributions.
The wave functions of a photon fluctuating into a dipole and of the dipole forming a vector meson are denoted by $\Psi_{\gamma^*}$ and $\Psi_{\rm V}$, respectively. The virtuality of the photon is $Q^2$, while $\vec{b}$ represents the impact parameter, and  $\vec{\Delta}$ is the momentum transferred in the interaction, with $\vec{\Delta}^2 \equiv-t$. The dipole has a transverse size  $\vec{r}$, and $z$ is the fraction of the photon momentum carried by the quark.
In this work, the targets H that we consider are proton (p) and lead (Pb). The centre-of-mass energy per nucleon of the photon--target system ($W$) is related to Bjorken-$x$ through 
\begin{equation}\label{eq:bjorkenx}
x = \frac{Q^2+M^2}{Q^2+W^2},
\end{equation}
with $M$ the mass of the vector meson. 
The factor $R_g ^{\rm T,L}$ is called the skewedness correction~\cite{Shuvaev:1999ce}, which is computed as
\begin{equation}
R^{\rm T,L}_g(\lambda^{\rm T,L}_g) = \frac{2^{2\lambda^{\rm T,L}_g+3}}{\sqrt{\pi}}\frac{\Gamma(\lambda^{\rm T,L}_g+5/2)}{\Gamma(\lambda^{\rm T,L}_g+4)},
\label{eq:Rg}
\end{equation}
with
\begin{equation}
\lambda^{\rm T,L}_g\equiv \frac{\partial\ln(A_{\rm T,L})}{\partial\ln(1/x)}
\label{eq:lambda}
\end{equation}
calculated at $t=0$.

 The cross section for a colour dipole scattering off a proton is
 \begin{equation}
\frac{\mathrm{d} \sigma^{\rm dip}_{\rm p}}{\mathrm{d}\vec{b}} = \sigma_0 N(x,r)T_{\rm p}(\vec{b}), 
\label{VM-dipole-cs}
\end{equation}
where the dipole scattering amplitude is given by the GBW model~\cite{GolecBiernat:1998js}
\begin{equation}
N(x,r) = \left[ 1 - \exp \left( -\frac{r^2 Q_s ^2 (x)}{4} \right) \right],
\label{VM-dipole-cs-GBW}
\end{equation}
with $Q_s(x)= Q_0 ^2 \left( x_0/x\right) ^{\lambda}$  the so-called  saturation scale. The dipole cross section off Pb targets reads
\begin{equation}
\left(\frac{{\rm d}\sigma^{\rm dip}_{\rm Pb}}{{\rm d}\vec{b}} \right)= 2\left[1-\left(1-\frac{1}{2A}\sigma_0 N(x,r)T_{\rm Pb}(\vec{b})\right)^A\right].
\label{eq:nuclear}
\end{equation}

The proton profile  $T_{\rm p} (\vec{b})$ is
\begin{equation}
T_{\rm p}(\vec{b}) = \frac{1}{N_{\rm hs}} \sum \limits_{i=1}^{N_{\rm hs}} T_{\rm hs} \left( \vec{b} - \vec{b}_i \right),
\label{VM-hs-T}
\end{equation}
where the hot-spot profile is 
\begin{equation}
T_{\rm hs} (\vec{b} - \vec{b}_i) = \frac{1}{2\pi B_{\rm hs}} \exp \left( -\frac{\left( \vec{b} - \vec{b}_i\right)^2}{2B_{\rm hs}} \right).
\label{VM-hs}
\end{equation}

The position of the hot spots (hs), $\vec{b_i}$, is sampled from a two-dimensional Gaussian distribution of width $B_{\rm p}$ and centred at (0,0). Thus, the parameters $B_{\rm p}$ and  $B_{\rm hs}$ represent one-half of the averaged squared radius of the proton and of the hot spot, respectively.
With this interpretation, $\sigma_0 = 4\pi B_{\rm p}$ is twice the overall transverse area of the proton.

The key feature of our model is the evolution of the number of hot spots ($N_{\rm hs}$) with energy, in order to reflect the raise of the gluon distribution, as Bjorken-$x$ decreases, observed in HERA data~\cite{H1:2015ubc}. 
$N_{\rm hs}$ is
sampled from a zero-truncated Poisson distribution, where the Poisson distribution has a mean value
\begin{equation}
\langle N_{ hs}(x) \rangle = p_0x^{p_1}(1+p_2\sqrt{x}).
\label{eq:Nhsx}
\end{equation}
The form of the parameterisation is inspired by the initial conditions used when fitting PDFs. 
The nuclear profile takes a similar form
\begin{equation}
T_{\rm Pb}(\vec{b}) =  \frac{1}{2\pi B_{\rm p}}  \sum \limits_{j=1}^{A=208} \exp \left( -\frac{\left( \vec{b} - \vec{b}_j\right)^2}{2B_{\rm p}} \right).\label{VM-Pb-P}
\end{equation}
where the index $j$ represents the nucleons in Pb, and their positions are sampled from a nuclear thickness function utilising a Woods-Saxon distribution. In this case, the hot-spot profile is 
\begin{equation}
T_{\rm hs} (\vec{b} - \vec{b}_i) = \frac{1}{2\pi B_{\rm hs}} \sum \limits_{i=1}^{A=208}
\frac{1}{N_{\rm hs}} \sum \limits_{j=1}^{N_{\rm hs}} 
 \exp \left( -\frac{\left( \vec{b} - \vec{b}_i- \vec{b}_j\right)^2}{2B_{\rm hs}} \right).
\label{VM-hs-Pb}
\end{equation}

For the vector meson wave function, we use boosted Gaussian model \cite{Nemchik:1994fp, Nemchik:1996cw, Forshaw:2003ki} with parameter values listed in Table I of Ref.\cite{Bendova:2018bbb}.
The values of the parameters of our model are discussed in Sec.~\ref{sec:Discussion}.

\section{The onset of gluon saturation in \gPb collisions
\label{sec:Results}}

\begin{figure}[!t]
\centering 
\includegraphics[width=0.48\textwidth]{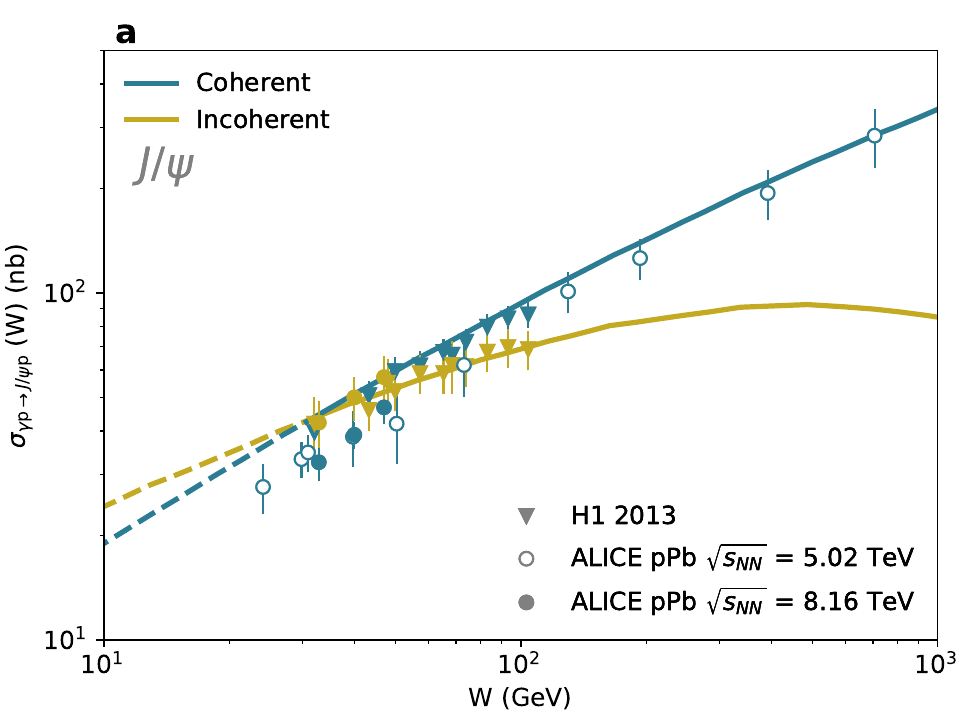}
\includegraphics[width=0.48\textwidth]{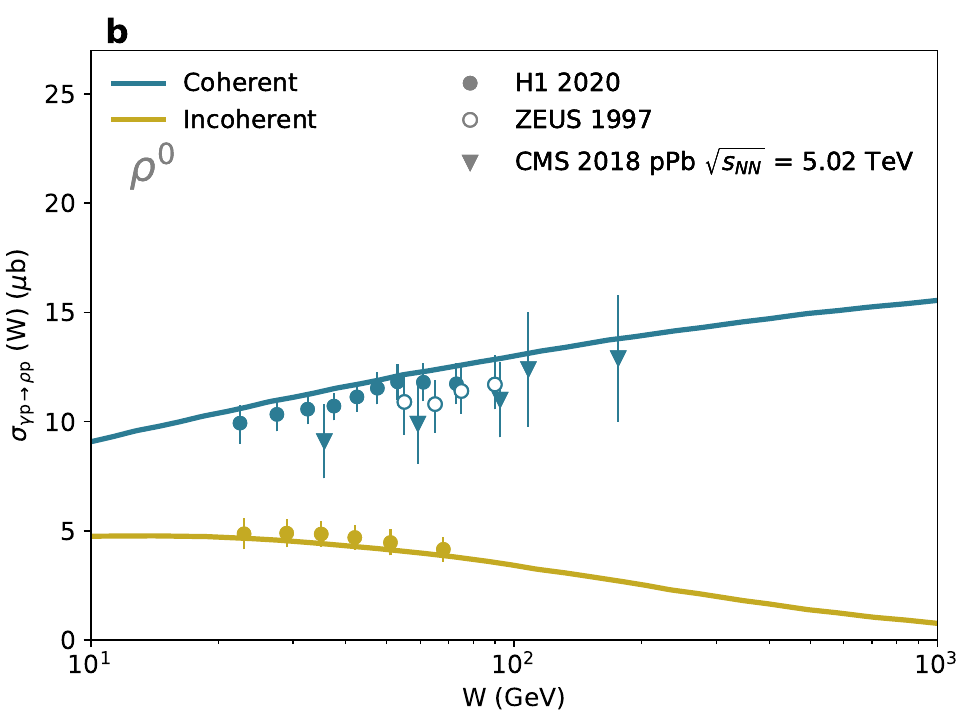}
\caption{Diffractive photo-production of \Jpsi (\textbf{a}) and \Rz (\textbf{b}) off protons for the coherent (blue) and incoherent (gold) processes.  The markers show measured data from the H1~\cite{H1:2013okq,H1:2020lzc}, ALICE~\cite{ALICE:2014eof,ALICE:2018oyo,ALICE:2023mfc}, and CMS~\cite{CMS:2019awk} collaborations, while the lines depict the predictions of our model. The dashed line represents values of $W$ that correspond to $x$ greater than 0.01, where the validity of the formalism is questionable.
\label{fig:gp} 
}
\end{figure}

As a benchmark, in Fig.~\ref{fig:gp}, we present the predictions of our model for $\gamma{\rm p}$ collisions and compare them to HERA~\cite{H1:2013okq,H1:2020lzc} and LHC~\cite{ALICE:2014eof,ALICE:2018oyo,ALICE:2023mfc,CMS:2019awk} data. Figure~\ref{fig:gp}~(a) shows the coherent and incoherent cross sections (known at HERA as  {\em exclusive} and {\em dissociative}, respectively)  for the diffractive photo-production of a \Jpsi vector meson. The coherent cross section rises with energy as a power law as does the prediction of our model. The available data for the incoherent process also rises with energy, but in this case, the model predicts that at around 500 GeV---outside the range of HERA, but inside that of the LHC---the cross section starts to decrease. The fact that the variance decreases, see Eq.~(\ref{VM-cs-diff-disoc}), signifies that the configurations start to resemble each other, which marks the onset of saturation. This behaviour was already discussed in our earlier publications~\cite{Cepila:2016uku,Cepila:2018zky,Bendova:2018bbb}, but at that time there was no data above the region where the incoherent cross section starts to decrease. 

Recently, the H1 collaboration has published new results on diffractive photo-production of a \Rz vector meson~\cite{H1:2020lzc}. Figure~\ref{fig:gp} (b) shows the new data and compares them to the predictions of our model. In this case, the coherent cross section rises with energy, but the incoherent cross section decreases, as we predicted already in Ref.~\cite{Cepila:2018zky}. This behaviour is striking, but the low value of the \Rz mass could cast a doubt on the applicability of a perturbative QCD approach. Nonetheless, the experimental confirmation of this prediction gives us confidence in the broad qualitative behaviour of our model.  Moreover, the larger mass of the \Jpsi meson, which is the focus of this work, is large enough to justify the applicability of perturbative QCD.

Coherent and incoherent diffractive production off nuclear targets offers the advantage that saturation sets in at a lower energy than for the case of proton targets  due to their larger saturation scale; see e.g. Ref.~\cite{Accardi:2012qut}. In this Letter, we utilise another characteristic of nuclei, namely the existence of different size scales in the transverse distribution of nuclear matter. Coherent processes sample the average of the colour fields of the full nuclei, so they are sensitive to nuclear sizes, $\mathcal{O}(10\, {\rm fm})$. Incoherent processes are sensitive to {\em two} different size scales, that of nucleons, $\mathcal{O}(1\, {\rm fm})$), and that of hot spots  ($\mathcal{O}(0.1\, {\rm fm})$).  It is expected that saturation are mainly linked to the hot-spot degrees of freedom, so we propose to study the energy dependence of incoherent photo-production of vector mesons in diffractive processes at different values of the Mandelstam-$t$ variable. Lower values of \mant are dominated by the contribution of large size scales; conversely, the cross section at large values of \mant is determined mainly by the variance of objects with a small transverse size, which in our model are the hot spots.

 \begin{figure}[!t]
\centering 
\includegraphics[width=0.48\textwidth]{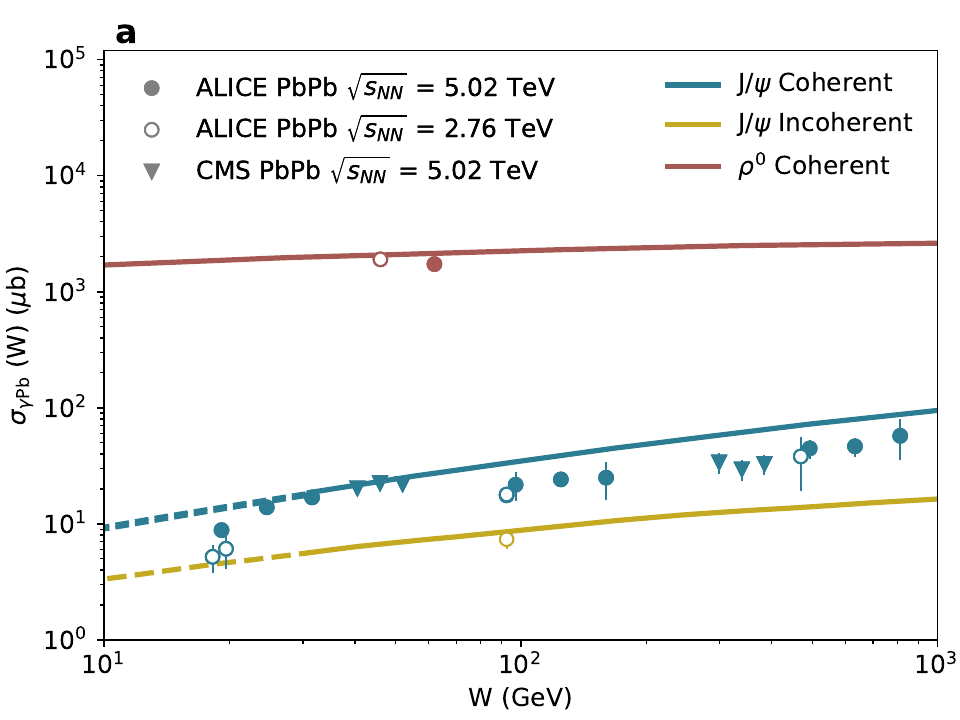}
\includegraphics[width=0.48\textwidth]{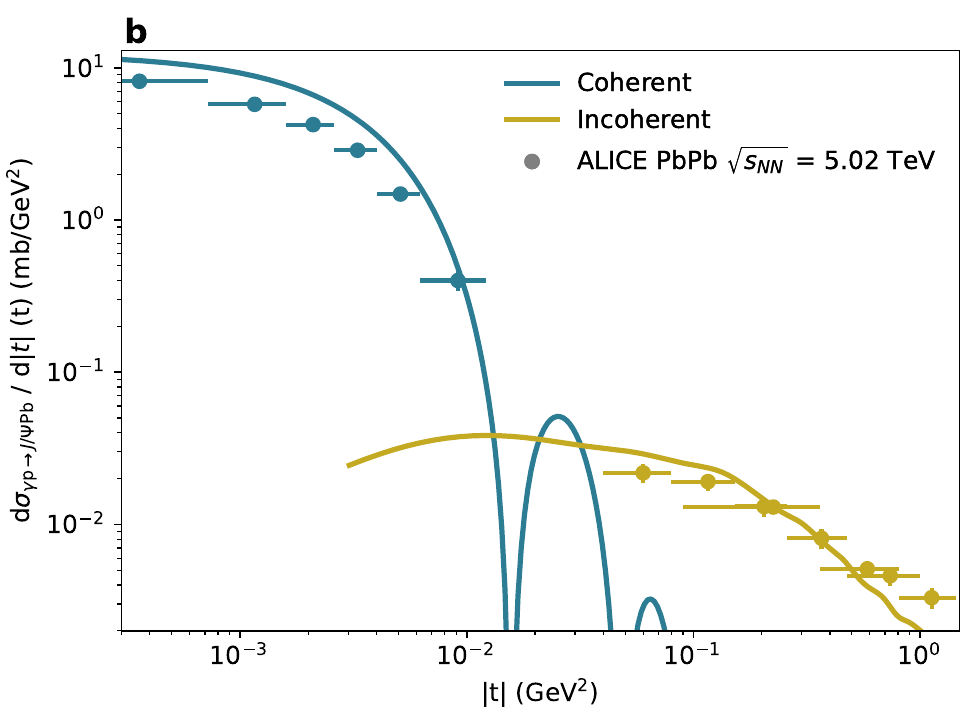}
\caption{\textbf{a:} Energy dependence of \Rz and \Jpsi photo-production off Pb. \textbf{b:} Mandelstam-$t$ dependence of coherent (blue) and incoherent (gold) \Jpsi photo-production off Pb at an energy $W\approx125$~GeV. The markers show data from the ALICE~\cite{ALICE:2015nbw,ALICE:2020ugp,ALICE:2021tyx,ALICE:2023gcs,ALICE:2023jgu} and CMS~\cite{CMS:2023snh} collaborations at the LHC, while the lines depict the predictions of our model.
\label{fig:gPb} 
}
\end{figure}

Figure~\ref{fig:gPb}  benchmarks our model against the existing LHC data~\cite{ALICE:2021tyx,ALICE:2023gcs,CMS:2023snh,ALICE:2023jgu} for the diffractive photo-production of \Rz and \Jpsi. Figure~\ref{fig:gPb}~(a) shows that our model is able to give a good reproduction of the energy dependence of coherent production across the almost two orders of magnitude covered by the \Jpsi data and that it also describes correctly the \Rz production. Figure~\ref{fig:gPb}~(b) shows the Mandelstam-$t$ dependence for both the coherent and the incoherent \Jpsi production. Our model describes correctly the behaviour of the incoherent cross section within the current experimental errors, but it shows a steeper slope than the data on coherent production. Thus, the latter data seem to indicate an effective slightly larger nucleus at these energies that assumed in the model. There is not enough experimental data to reliable model such an energy dependence of the nuclear size. Nonetheless the effect is small, and would not affect the qualitatively conclusions of our work. The cross section for incoherent J$/\psi$ production from Ref.~\cite{ALICE:2013wjo}, was scaled by the photon flux at midrapidity taken from Ref.~\cite{Contreras:2016pkc}. As the figure shows, at the small \mant region the cross section is dominated by the coherent contribution, while at larger \mant the incoherent process dominates, as explained above. 

\begin{figure}[!t]
\centering 
\includegraphics[width=0.48\textwidth]{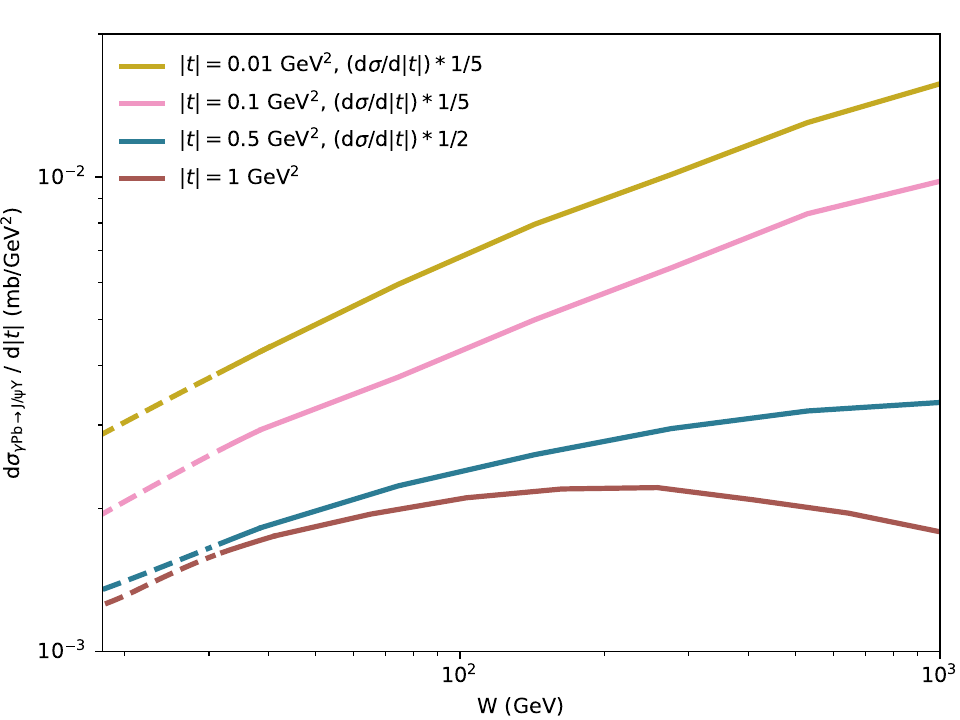}
\caption{Prediction of the energy-dependent hot-spot model for the incoherent photo-production of \Jpsi vector mesons off Pb in diffractive interactions. The lines depict the energy dependence of this process at different values of the Mandelstam-$t$ variable. Some of the lines have  been scaled to improve the readability of the figure.
\label{fig:gPbt} 
}
\end{figure}

Figure~\ref{fig:gPbt} shows the main result of this Letter: the cross section for the incoherent photo-production of \Jpsi vector mesons off Pb in diffractive interactions. The energy dependence at several values of \mant is shown. For small \mant values, the cross section raises with energy, but at larger values of \mant, where the cross section is dominated by the hot-spot contribution, the rise of the cross section reaches a maximum at around a few hundred GeV and then decreases. The maximum marks the onset of saturation effects and it is well within the reach of the LHC.

\section{Discussion \label{sec:Discussion}}

With respect to our previous work, there are two changes in the formalism. The first refers to Eq.~(\ref{VM-amplitude}), where now we use the  prescription of Ref.~\cite{Hatta:2017cte} for the exponent.  This amounts to use $((1/2)-z)$ instead of $(1-z)$ as it was used before~\cite{Kowalski:2006hc}. This change has almost no effect on the predictions for the coherent processes, but it changes the predictions for the incoherent case. The previous good
description of available data is recovered by using new values of the $p_i$ parameters, see Eq.~(\ref{eq:Nhsx}). After these changes, the results are equivalent to those that we have reported before. The values of the parameters of our model that control the behaviour of the hot spots are  then $B_{\rm p} = 4.75$ GeV$^{-2}$, $B_{\rm hs} = 0.8$ GeV$^{-2}$, $p_0 = 0.015$, $p_1 = -0.58$, $p_2 = 300$. 
As discussed in greater detail in our earlier work~\cite{Cepila:2016uku}, the value of $Q_0$ was set to 1 GeV, $x_0$ was set to $2 \cdot 10 ^{-4}$, and $\lambda = 0.21$; where the latter was chosen to describe the behaviour of $F_2$ at a scale $Q^2=2.7$ GeV$^2$ corresponding to that of the J$/\psi$ vector meson in photoproduction.

When considering the case of \Rz photo-production we set $B_{\rm p}=6.4$ GeV$^{-2}$ to be consistent with the findings by the H1 collaboration~\cite{Aaron:2009xp}. It is worth noting that the change of the value of this parameter is just a phenomenological tool to extend the model to describe also the photoproduction of  $\rho^0$, where there is no clear perturbative scale. The photon viruality is set to $Q^2=$ 0.05\,GeV$^2$ in all calculations.
The total cross section is determined by integrating the differential cross section, see Eq. (\ref{VM-cs-diff-excl}, \ref{VM-cs-diff-disoc}), over \mant. For the case of the proton as a target the integration is performed within the range specified by the available experimental data. In the case of photo-production of $\Jpsi$ off protons the used ranges are $\mant < 1.2 \textrm{ GeV}^2$ for exclusive and $\mant < 8 \textrm{ GeV}^2$ for dissociative process ~\cite{H1:2013okq}. For exclusive and dissociative $\Rz$ photo-production, the corresponding integration cuts are $\mant < 0.5 \textrm{ GeV}^2$ ~\cite{CMS:2019awk} and $\mant < 1.5 \textrm{ GeV}^2$~\cite{H1:2020lzc} respectively. The measurements of the energy dependence of $\Rz$ and $\Jpsi$ photo-production off Pb have no restriction on \mant.

Another change with respect to our previous work refers to Eq.~(\ref{eq:nuclear}). In the past, we used the same formalism but took the limit of a large $A$. In this case, we did not take the limit in preparation for predictions of the incoming oxygen-oxygen collisions planned for the LHC~\cite{ALICE-PUBLIC-2021-004}. In the case of oxygen, the large $A$ limit is not justified. For Pb, Eq.~(\ref{eq:nuclear}) produces results numerically very similar to the ones coming from the large $A$ limit.

The ALICE and CMS collaborations have recently published data on coherent \Jpsi production reaching $W$ values slightly above 800 GeV~\cite{ALICE:2023jgu,CMS:2023snh}. As the \Jpsi signature is the same for coherent and incoherent processes, and  as both collaborations are equipped with zero-degree calorimetres, they could perform the measurement predicted in Fig.~\ref{fig:gPbt} in a similar energy range by tagging neutrons at beam rapidity~\cite{Strikman:2005ze,Kryshen:2023bxy}

The shape for the $W$ dependence at a fixed value of $|t|$ predicted
by our model and shown in Fig.~\ref{fig:gPbt} can be  described by 
\begin{equation}
f(W) = N\, (W/W_0)^{\delta} \exp{\big(-(W/W_0)(\delta/W_{\rm max})\big)}.
\end{equation}
Fitting this function to the predictions at $|t| = 0.1$ GeV$^2$ and
$|t| = 1.0$ GeV$^2$ we find $W_{\rm max}$ to be $1550\pm107$ GeV and
$297\pm6$ GeV, respectively. In these fits a fixed value of $W_0=90$
GeV was used. For completeness we report the values found for the parameter $\delta$: $0.45\pm0.01$ and $0.26\pm0.01$, respectively.  Note that Eq.~(\ref{eq:bjorkenx}) is an approximation, with the
full equation being $x = (Q^2+M^2-t)/(Q^2+W^2)$. If the full
expression is used, the quoted $W_{\rm max}$ values move to $1555\pm107$ GeV 
and $312\pm7$ GeV, respectively. The values of  $\delta$ do not change.

Looking at Fig.~\ref{fig:gPbt} it is tempting to go to higher values of $|t|$. Numerically, our model can produce such predictions but two issues have to be taken into account: (a) the larger the value of $|t|$ the farther away one is from the forward limit used to set up the model, (b) furthermore, once $|t|$ is large enough, another perturbative scale enters the problem and it has to be taken into account properly. Fortunately, such very large values of $|t|$ are not optimal to sample the hot-spot sizes of our model, so in the context of this study it is not needed to go higher than 1 GeV in $|t|$.

\section{Summary and outlook
\label{sec:Summary}}

Utilising the energy-dependent hot-spot model, properly bench-marked with HERA and LHC data, we propose to measure the onset of saturation by the energy dependence of the cross section for incoherent \Jpsi photo-production at large Mandelstam-$t$, the region dominated by the hot-spot contribution. We predict that the onset of saturation can be measured using ultra-peripheral Pb--Pb collisions at the LHC.

The measurement could be performed with data recorded during the LHC Run 2 (2015-2018). Furthermore, in 2023  the LHC delivered the first Pb–Pb data sample of the LHC Run 3, and more data is expected in the following years during the LHC Run 3 and 4. In total, one expects to collect on the order of one million \Jpsi in the $\mu^+\mu^-$ channel~\cite{Citron:2018lsq}. This amount of expected events should allow to perform  the measurement of this process with just a few percent uncertainty. 
\section*{Acknowledgements}
This work was partially funded by the Czech Science Foundation (GAČR), project No. 22-27262S.

\bibliography{bibliography}

\begin{thebibliography}{10}
\expandafter\ifx\csname url\endcsname\relax
  \def\url#1{\texttt{#1}}\fi
\expandafter\ifx\csname urlprefix\endcsname\relax\def\urlprefix{URL }\fi
\expandafter\ifx\csname href\endcsname\relax
  \def\href#1#2{#2} \def\path#1{#1}\fi

\bibitem{Gribov:1983ivg}
L.~V. Gribov, E.~M. Levin, M.~G. Ryskin, {Semihard Processes in QCD}, Phys. Rept. 100 (1983) 1--150.
\newblock \href {https://doi.org/10.1016/0370-1573(83)90022-4} {\path{doi:10.1016/0370-1573(83)90022-4}}.

\bibitem{Mueller:1989st}
A.~H. Mueller, {Small x Behavior and Parton Saturation: A QCD Model}, Nucl. Phys. B 335 (1990) 115--137.
\newblock \href {https://doi.org/10.1016/0550-3213(90)90173-B} {\path{doi:10.1016/0550-3213(90)90173-B}}.

\bibitem{Albacete:2014fwa}
J.~L. Albacete, C.~Marquet, {Gluon saturation and initial conditions for relativistic heavy ion collisions}, Prog. Part. Nucl. Phys. 76 (2014) 1--42.
\newblock \href {http://arxiv.org/abs/1401.4866} {\path{arXiv:1401.4866}}, \href {https://doi.org/10.1016/j.ppnp.2014.01.004} {\path{doi:10.1016/j.ppnp.2014.01.004}}.

\bibitem{H1:2015ubc}
H.~Abramowicz, et~al., {Combination of measurements of inclusive deep inelastic ${e^{\pm }p}$ scattering cross sections and QCD analysis of HERA data}, Eur. Phys. J. C 75~(12) (2015) 580.
\newblock \href {http://arxiv.org/abs/1506.06042} {\path{arXiv:1506.06042}}, \href {https://doi.org/10.1140/epjc/s10052-015-3710-4} {\path{doi:10.1140/epjc/s10052-015-3710-4}}.

\bibitem{Ryskin:1992ui}
M.~G. Ryskin, {Diffractive J / psi electroproduction in LLA QCD}, Z. Phys. C 57 (1993) 89--92.
\newblock \href {https://doi.org/10.1007/BF01555742} {\path{doi:10.1007/BF01555742}}.

\bibitem{Kowalski:2006hc}
H.~Kowalski, L.~Motyka, G.~Watt, {Exclusive diffractive processes at HERA within the dipole picture}, Phys. Rev. D74 (2006) 074016.
\newblock \href {http://arxiv.org/abs/hep-ph/0606272} {\path{arXiv:hep-ph/0606272}}, \href {https://doi.org/10.1103/PhysRevD.74.074016} {\path{doi:10.1103/PhysRevD.74.074016}}.

\bibitem{Newman:2013ada}
P.~Newman, M.~Wing, {The Hadronic Final State at HERA}, Rev. Mod. Phys. 86~(3) (2014) 1037.
\newblock \href {http://arxiv.org/abs/1308.3368} {\path{arXiv:1308.3368}}, \href {https://doi.org/10.1103/RevModPhys.86.1037} {\path{doi:10.1103/RevModPhys.86.1037}}.

\bibitem{Contreras:2015dqa}
J.~G. Contreras, J.~D. Tapia~Takaki, {Ultra-peripheral heavy-ion collisions at the LHC}, Int. J. Mod. Phys. A 30 (2015) 1542012.
\newblock \href {https://doi.org/10.1142/S0217751X15420129} {\path{doi:10.1142/S0217751X15420129}}.

\bibitem{Klein:2019qfb}
S.~R. Klein, H.~M\"antysaari, {Imaging the nucleus with high-energy photons}, Nature Rev. Phys. 1~(11) (2019) 662--674.
\newblock \href {http://arxiv.org/abs/1910.10858} {\path{arXiv:1910.10858}}, \href {https://doi.org/10.1038/s42254-019-0107-6} {\path{doi:10.1038/s42254-019-0107-6}}.

\bibitem{Accardi:2012qut}
A.~Accardi, et~al., {Electron Ion Collider: The Next QCD Frontier}: {Understanding the glue that binds us all}, Eur. Phys. J. A 52~(9) (2016) 268.
\newblock \href {http://arxiv.org/abs/1212.1701} {\path{arXiv:1212.1701}}, \href {https://doi.org/10.1140/epja/i2016-16268-9} {\path{doi:10.1140/epja/i2016-16268-9}}.

\bibitem{LHeCStudyGroup:2012zhm}
J.~L. Abelleira~Fernandez, et~al., {A Large Hadron Electron Collider at CERN: Report on the Physics and Design Concepts for Machine and Detector}, J. Phys. G 39 (2012) 075001.
\newblock \href {http://arxiv.org/abs/1206.2913} {\path{arXiv:1206.2913}}, \href {https://doi.org/10.1088/0954-3899/39/7/075001} {\path{doi:10.1088/0954-3899/39/7/075001}}.

\bibitem{Good:1960ba}
M.~L. Good, W.~D. Walker, {Diffraction disssociation of beam particles}, Phys. Rev. 120 (1960) 1857--1860.
\newblock \href {https://doi.org/10.1103/PhysRev.120.1857} {\path{doi:10.1103/PhysRev.120.1857}}.

\bibitem{Miettinen:1978jb}
H.~I. Miettinen, J.~Pumplin, {Diffraction Scattering and the Parton Structure of Hadrons}, Phys. Rev. D 18 (1978) 1696.
\newblock \href {https://doi.org/10.1103/PhysRevD.18.1696} {\path{doi:10.1103/PhysRevD.18.1696}}.

\bibitem{Mantysaari:2016ykx}
H.~M\"antysaari, B.~Schenke, {Evidence of strong proton shape fluctuations from incoherent diffraction}, Phys. Rev. Lett. 117~(5) (2016) 052301.
\newblock \href {http://arxiv.org/abs/1603.04349} {\path{arXiv:1603.04349}}, \href {https://doi.org/10.1103/PhysRevLett.117.052301} {\path{doi:10.1103/PhysRevLett.117.052301}}.

\bibitem{Traini:2018hxd}
M.~C. Traini, J.-P. Blaizot, {Diffractive incoherent vector meson production off protons: a quark model approach to gluon fluctuation effects}, Eur. Phys. J. C 79~(4) (2019) 327.
\newblock \href {http://arxiv.org/abs/1804.06110} {\path{arXiv:1804.06110}}, \href {https://doi.org/10.1140/epjc/s10052-019-6826-0} {\path{doi:10.1140/epjc/s10052-019-6826-0}}.

\bibitem{Kumar:2021zbn}
A.~Kumar, T.~Toll, {Investigating the structure of gluon fluctuations in the proton with incoherent diffraction at HERA}, Eur. Phys. J. C 82~(9) (2022) 837.
\newblock \href {http://arxiv.org/abs/2106.12855} {\path{arXiv:2106.12855}}, \href {https://doi.org/10.1140/epjc/s10052-022-10774-3} {\path{doi:10.1140/epjc/s10052-022-10774-3}}.

\bibitem{Demirci:2022wuy}
S.~Demirci, T.~Lappi, S.~Schlichting, {Proton hot spots and exclusive vector meson production}, Phys. Rev. D 106~(7) (2022) 074025.
\newblock \href {http://arxiv.org/abs/2206.05207} {\path{arXiv:2206.05207}}, \href {https://doi.org/10.1103/PhysRevD.106.074025} {\path{doi:10.1103/PhysRevD.106.074025}}.

\bibitem{Mantysaari:2020axf}
H.~M\"antysaari, {Review of proton and nuclear shape fluctuations at high energy}, Rept. Prog. Phys. 83~(8) (2020) 082201.
\newblock \href {http://arxiv.org/abs/2001.10705} {\path{arXiv:2001.10705}}, \href {https://doi.org/10.1088/1361-6633/aba347} {\path{doi:10.1088/1361-6633/aba347}}.

\bibitem{Cepila:2016uku}
J.~Cepila, J.~G. Contreras, J.~D. Tapia~Takaki, {Energy dependence of dissociative $\mathrm{J/}\psi$ photoproduction as a signature of gluon saturation at the LHC}, Phys. Lett. B766 (2017) 186--191.
\newblock \href {http://arxiv.org/abs/1608.07559} {\path{arXiv:1608.07559}}, \href {https://doi.org/10.1016/j.physletb.2016.12.063} {\path{doi:10.1016/j.physletb.2016.12.063}}.

\bibitem{Armesto:1996kt}
N.~Armesto, M.~A. Braun, E.~G. Ferreiro, C.~Pajares, {Percolation approach to quark - gluon plasma and J / psi suppression}, Phys. Rev. Lett. 77 (1996) 3736--3738.
\newblock \href {http://arxiv.org/abs/hep-ph/9607239} {\path{arXiv:hep-ph/9607239}}, \href {https://doi.org/10.1103/PhysRevLett.77.3736} {\path{doi:10.1103/PhysRevLett.77.3736}}.

\bibitem{Cepila:2017nef}
J.~Cepila, J.~G. Contreras, M.~Krelina, {Coherent and incoherent $\mathrm{J/}\psi$ photonuclear production in an energy-dependent hot-spot model}, Phys. Rev. C97~(2) (2018) 024901.
\newblock \href {http://arxiv.org/abs/1711.01855} {\path{arXiv:1711.01855}}, \href {https://doi.org/10.1103/PhysRevC.97.024901} {\path{doi:10.1103/PhysRevC.97.024901}}.

\bibitem{Cepila:2018zky}
J.~Cepila, J.~G. Contreras, M.~Krelina, J.~D. Tapia~Takaki, {Mass dependence of vector meson photoproduction off protons and nuclei within the energy-dependent hot-spot model}, Nucl. Phys. B934 (2018) 330--340.
\newblock \href {http://arxiv.org/abs/1804.05508} {\path{arXiv:1804.05508}}, \href {https://doi.org/10.1016/j.nuclphysb.2018.07.010} {\path{doi:10.1016/j.nuclphysb.2018.07.010}}.

\bibitem{Bendova:2018bbb}
D.~Bendova, J.~Cepila, J.~G. Contreras, {Dissociative production of vector mesons at electron-ion colliders}, Phys. Rev. D 99~(3) (2019) 034025.
\newblock \href {http://arxiv.org/abs/1811.06479} {\path{arXiv:1811.06479}}, \href {https://doi.org/10.1103/PhysRevD.99.034025} {\path{doi:10.1103/PhysRevD.99.034025}}.

\bibitem{Krelina:2019gee}
M.~Krelina, V.~P. Goncalves, J.~Cepila, {Coherent and incoherent vector meson electroproduction in the future electron-ion colliders: the hot-spot predictions}, Nucl. Phys. A 989 (2019) 187--200.
\newblock \href {http://arxiv.org/abs/1905.06759} {\path{arXiv:1905.06759}}, \href {https://doi.org/10.1016/j.nuclphysa.2019.06.009} {\path{doi:10.1016/j.nuclphysa.2019.06.009}}.

\bibitem{Shuvaev:1999ce}
A.~G. Shuvaev, K.~J. Golec-Biernat, A.~D. Martin, M.~G. Ryskin, {Off diagonal distributions fixed by diagonal partons at small x and xi}, Phys. Rev. D60 (1999) 014015.
\newblock \href {http://arxiv.org/abs/hep-ph/9902410} {\path{arXiv:hep-ph/9902410}}, \href {https://doi.org/10.1103/PhysRevD.60.014015} {\path{doi:10.1103/PhysRevD.60.014015}}.

\bibitem{GolecBiernat:1998js}
K.~J. Golec-Biernat, M.~Wusthoff, {Saturation effects in deep inelastic scattering at low Q**2 and its implications on diffraction}, Phys. Rev. D59 (1998) 014017.
\newblock \href {http://arxiv.org/abs/hep-ph/9807513} {\path{arXiv:hep-ph/9807513}}, \href {https://doi.org/10.1103/PhysRevD.59.014017} {\path{doi:10.1103/PhysRevD.59.014017}}.

\bibitem{Nemchik:1994fp}
J.~Nemchik, N.~N. Nikolaev, B.~G. Zakharov, {Scanning the BFKL pomeron in elastic production of vector mesons at HERA}, Phys. Lett. B 341 (1994) 228--237.
\newblock \href {http://arxiv.org/abs/hep-ph/9405355} {\path{arXiv:hep-ph/9405355}}, \href {https://doi.org/10.1016/0370-2693(94)90314-X} {\path{doi:10.1016/0370-2693(94)90314-X}}.

\bibitem{Nemchik:1996cw}
J.~Nemchik, N.~N. Nikolaev, E.~Predazzi, B.~G. Zakharov, {Color dipole phenomenology of diffractive electroproduction of light vector mesons at HERA}, Z. Phys. C 75 (1997) 71--87.
\newblock \href {http://arxiv.org/abs/hep-ph/9605231} {\path{arXiv:hep-ph/9605231}}, \href {https://doi.org/10.1007/s002880050448} {\path{doi:10.1007/s002880050448}}.

\bibitem{Forshaw:2003ki}
J.~R. Forshaw, R.~Sandapen, G.~Shaw, {Color dipoles and rho, phi electroproduction}, Phys. Rev. D 69 (2004) 094013.
\newblock \href {http://arxiv.org/abs/hep-ph/0312172} {\path{arXiv:hep-ph/0312172}}, \href {https://doi.org/10.1103/PhysRevD.69.094013} {\path{doi:10.1103/PhysRevD.69.094013}}.

\bibitem{H1:2013okq}
C.~Alexa, et~al., {Elastic and Proton-Dissociative Photoproduction of J/psi Mesons at HERA}, Eur. Phys. J. C 73~(6) (2013) 2466.
\newblock \href {http://arxiv.org/abs/1304.5162} {\path{arXiv:1304.5162}}, \href {https://doi.org/10.1140/epjc/s10052-013-2466-y} {\path{doi:10.1140/epjc/s10052-013-2466-y}}.

\bibitem{H1:2020lzc}
V.~Andreev, et~al., {Measurement of Exclusive $\pi^{+}\pi^{-}$ and $\rho^0$ Meson Photoproduction at HERA}, Eur. Phys. J. C 80~(12) (2020) 1189.
\newblock \href {http://arxiv.org/abs/2005.14471} {\path{arXiv:2005.14471}}, \href {https://doi.org/10.1140/epjc/s10052-020-08587-3} {\path{doi:10.1140/epjc/s10052-020-08587-3}}.

\bibitem{ALICE:2014eof}
B.~B. Abelev, et~al., {Exclusive $\mathrm{J/}\psi$ photoproduction off protons in ultra-peripheral p-Pb collisions at $\sqrt{s_{\rm NN}}=5.02$ TeV}, Phys. Rev. Lett. 113~(23) (2014) 232504.
\newblock \href {http://arxiv.org/abs/1406.7819} {\path{arXiv:1406.7819}}, \href {https://doi.org/10.1103/PhysRevLett.113.232504} {\path{doi:10.1103/PhysRevLett.113.232504}}.

\bibitem{ALICE:2018oyo}
S.~Acharya, et~al., {Energy dependence of exclusive $\mathrm {J}/\psi $ photoproduction off protons in ultra-peripheral p\textendash{}Pb collisions at $\sqrt{s_{\mathrm {\scriptscriptstyle NN}}} = 5.02$ TeV}, Eur. Phys. J. C 79~(5) (2019) 402.
\newblock \href {http://arxiv.org/abs/1809.03235} {\path{arXiv:1809.03235}}, \href {https://doi.org/10.1140/epjc/s10052-019-6816-2} {\path{doi:10.1140/epjc/s10052-019-6816-2}}.

\bibitem{ALICE:2023mfc}
{Exclusive and dissociative J/$\psi$ photoproduction, and exclusive dimuon production, in p$-$Pb collisions at $\sqrt{s_{\rm NN}} = 8.16$ TeV} (4 2023).
\newblock \href {http://arxiv.org/abs/2304.12403} {\path{arXiv:2304.12403}}.

\bibitem{CMS:2019awk}
A.~M. Sirunyan, et~al., {Measurement of exclusive $\rho(770)^0$ photoproduction in ultraperipheral pPb collisions at $\sqrt{s_\mathrm{NN}} =$ 5.02 TeV}, Eur. Phys. J. C 79~(8) (2019) 702.
\newblock \href {http://arxiv.org/abs/1902.01339} {\path{arXiv:1902.01339}}, \href {https://doi.org/10.1140/epjc/s10052-019-7202-9} {\path{doi:10.1140/epjc/s10052-019-7202-9}}.

\bibitem{ALICE:2015nbw}
J.~Adam, et~al., {Coherent \ensuremath{\rho}$^{0}$ photoproduction in ultra-peripheral Pb--Pb collisions at $\sqrt{s_{\rm{NN}}} = 2.76$~TeV}, JHEP 09 (2015) 095.
\newblock \href {http://arxiv.org/abs/1503.09177} {\path{arXiv:1503.09177}}, \href {https://doi.org/10.1007/JHEP09(2015)095} {\path{doi:10.1007/JHEP09(2015)095}}.

\bibitem{ALICE:2020ugp}
S.~Acharya, et~al., {Coherent photoproduction of $\rho^{0}$ vector mesons in ultra-peripheral Pb--Pb collisions at $\sqrt{s_{\rm{NN}}} = 5.02$~TeV}, JHEP 06 (2020) 035.
\newblock \href {http://arxiv.org/abs/2002.10897} {\path{arXiv:2002.10897}}, \href {https://doi.org/10.1007/JHEP06(2020)035} {\path{doi:10.1007/JHEP06(2020)035}}.

\bibitem{ALICE:2021tyx}
S.~Acharya, et~al., {First measurement of the $|t|$-dependence of coherent $\textrm{J}/\psi$ photonuclear production}, Phys. Lett. B 817 (2021) 136280.
\newblock \href {http://arxiv.org/abs/2101.04623} {\path{arXiv:2101.04623}}, \href {https://doi.org/10.1016/j.physletb.2021.136280} {\path{doi:10.1016/j.physletb.2021.136280}}.

\bibitem{ALICE:2023gcs}
S.~Acharya, et~al., {First measurement of the $|t|$-dependence of incoherent J/$\psi$ photonuclear production} (5 2023).
\newblock \href {http://arxiv.org/abs/2305.06169} {\path{arXiv:2305.06169}}.

\bibitem{ALICE:2023jgu}
S.~Acharya, et~al., {Energy dependence of coherent photonuclear production of J/\ensuremath{\psi} mesons in ultra-peripheral Pb-Pb collisions at $ \sqrt{{\textrm{s}}_{\textrm{NN}}} $ = 5.02 TeV}, JHEP 10 (2023) 119.
\newblock \href {http://arxiv.org/abs/2305.19060} {\path{arXiv:2305.19060}}, \href {https://doi.org/10.1007/JHEP10(2023)119} {\path{doi:10.1007/JHEP10(2023)119}}.

\bibitem{CMS:2023snh}
A.~Tumasyan, et~al., {Probing small Bjorken-$x$ nuclear gluonic structure via coherent J/$\psi$ photoproduction in ultraperipheral PbPb collisions at $\sqrt{s_\mathrm{NN}}$ = 5.02 TeV} (3 2023).
\newblock \href {http://arxiv.org/abs/2303.16984} {\path{arXiv:2303.16984}}.

\bibitem{ALICE:2013wjo}
E.~Abbas, et~al., {Charmonium and $e^+e^-$ pair photoproduction at mid-rapidity in ultra-peripheral Pb-Pb collisions at $\sqrt{s_{\rm NN}}$=2.76 TeV}, Eur. Phys. J. C 73~(11) (2013) 2617.
\newblock \href {http://arxiv.org/abs/1305.1467} {\path{arXiv:1305.1467}}, \href {https://doi.org/10.1140/epjc/s10052-013-2617-1} {\path{doi:10.1140/epjc/s10052-013-2617-1}}.

\bibitem{Contreras:2016pkc}
J.~G. Contreras, {Gluon shadowing at small $x$ from coherent $\mathrm{J/}\psi$ photoproduction data at energies available at the CERN Large Hadron Collider}, Phys. Rev. C 96~(1) (2017) 015203.
\newblock \href {http://arxiv.org/abs/1610.03350} {\path{arXiv:1610.03350}}, \href {https://doi.org/10.1103/PhysRevC.96.015203} {\path{doi:10.1103/PhysRevC.96.015203}}.

\bibitem{Hatta:2017cte}
Y.~Hatta, B.-W. Xiao, F.~Yuan, {Gluon Tomography from Deeply Virtual Compton Scattering at Small-x}, Phys. Rev. D 95~(11) (2017) 114026.
\newblock \href {http://arxiv.org/abs/1703.02085} {\path{arXiv:1703.02085}}, \href {https://doi.org/10.1103/PhysRevD.95.114026} {\path{doi:10.1103/PhysRevD.95.114026}}.

\bibitem{Aaron:2009xp}
F.~D. Aaron, et~al., {Diffractive Electroproduction of rho and phi Mesons at HERA}, JHEP 05 (2010) 032.
\newblock \href {http://arxiv.org/abs/0910.5831} {\path{arXiv:0910.5831}}, \href {https://doi.org/10.1007/JHEP05(2010)032} {\path{doi:10.1007/JHEP05(2010)032}}.

\bibitem{ALICE-PUBLIC-2021-004}
\href{https://cds.cern.ch/record/2765973}{{ALICE physics projections for a short oxygen-beam run at the LHC}} (2021).
\newline\urlprefix\url{https://cds.cern.ch/record/2765973}

\bibitem{Strikman:2005ze}
M.~Strikman, M.~Tverskoy, M.~Zhalov, {Neutron tagging of quasielastic J/psi photoproduction off nucleus in ultraperipheral heavy ion collisions at RHIC energies}, Phys. Lett. B 626 (2005) 72--79.
\newblock \href {http://arxiv.org/abs/hep-ph/0505023} {\path{arXiv:hep-ph/0505023}}, \href {https://doi.org/10.1016/j.physletb.2005.08.083} {\path{doi:10.1016/j.physletb.2005.08.083}}.

\bibitem{Kryshen:2023bxy}
E.~Kryshen, M.~Strikman, M.~Zhalov, {Photoproduction of J/\ensuremath{\psi} with neutron tagging in ultraperipheral collisions of nuclei at RHIC and at the LHC}, Phys. Rev. C 108~(2) (2023) 024904.
\newblock \href {http://arxiv.org/abs/2303.12052} {\path{arXiv:2303.12052}}, \href {https://doi.org/10.1103/PhysRevC.108.024904} {\path{doi:10.1103/PhysRevC.108.024904}}.

\bibitem{Citron:2018lsq}
Z.~Citron, et~al., {Report from Working Group 5}: {Future physics opportunities for high-density QCD at the LHC with heavy-ion and proton beams}, CERN Yellow Rep. Monogr. 7 (2019) 1159--1410.
\newblock \href {http://arxiv.org/abs/1812.06772} {\path{arXiv:1812.06772}}, \href {https://doi.org/10.23731/CYRM-2019-007.1159} {\path{doi:10.23731/CYRM-2019-007.1159}}.

\end{thebibliography}

\end{document}